\documentclass[preprint,aps]{revtex4}
\newcommand{\be}{\begin{equation}}
\newcommand{\ee}{\end{equation}}
\newcommand{\ba}{\begin{eqnarray}}
\newcommand{\ea}{\end{eqnarray}}

\begin{document}

\draft

\title{Bonnor-type Black Dihole Solution\\
       in Brans-Dicke-Maxwell Theory}

\author{Hongsu Kim\footnote{e-mail : hongsu@astro.snu.ac.kr} and
Hyung Mok Lee\footnote{e-mail :hmlee@astro.snu.ac.kr}}

\address{Astronomy Program, SEES, Seoul National University, Seoul, 151-742, KOREA}

\begin{abstract}

It was originally thought that Bonnor's solution in Einstein-Maxwell theory describes 
a singular point-like magnetic dipole. Lately, however, it has been demonstrated that indeed
it may describe a black {\it dihole}, i.e., a pair of static, oppositely-charged 
extremal black holes with regular horizons. Motivated particularly by this new 
interpretation, in the present work, the construction and extensive analysis
of a solution in the context of the Brans-Dicke-Maxwell theory representing a black dihole
are attempted. It has been known for some time that the solution-generating algorithm of 
Singh and Rai produces stationary, axisymmetric, charged solutions in Brans-Dicke-Maxwell 
theory from the known such solutions in Einstein-Maxwell theory. Thus this algorithm of 
Singh and Rai's is employed in order to construct a Bonnor-type magnetic black dihole solution 
in Brans-Dicke-Maxwell theory from the known Bonnor solution in Einstein-Maxwell theory. 
The peculiar features of the new solution including internal infinity nature of the symmetry 
axis and its stability issue have been discussed in full detail.

\end{abstract}


\maketitle

\narrowtext

\newpage
\begin{center}
{\rm\bf I. Introduction}
\end{center}
 
Not surprisingly, exact solutions of general relativity describing {\it multiple}
black hole configurations are few since they would be highly non-trivial to find
or construct. Although one would naturally expect that such configurations might
in general involve a very complicated structure, there actually exist some simple
solutions having remarkable properties. These include the Majumdar-Papapetrou 
solutions \cite{papa} representing an arbitrary number of static, extremal charged
black holes, all with charges of the same sign and the multi-Schwarzschild solution
of Israel and Khan \cite{ik}. Besides, the C \cite{c} and the Ernst metrics \cite{ernst}
or the cosmological multi-black hole solutions of Kastor and Trachen \cite{kt},
all describing the black holes in relative motion, may also fall into this category of
multi-black hole solutions.   \\
In the present work, we are particularly interested in
Bonnor's solution. Several years ago, using a technique to generate a static,
axisymmetric solution of the Einstein-Maxwell theory from the stationary, axisymmetric
Kerr solution of the vacuum Einstein theory, Bonnor \cite{bonnor} constructed a
solution describing a magnetic dipole. Bonnor's solution was originally thought to 
describe a singular point-like dipole. Recently, the generalization of Bonnor's
solution in Einstein-Maxwell theory to its counterparts in Einstein-Maxwell-dilaton
theories with the correct analysis of the conical singularity structure of the
solutions was made by Davidson and Gedalin \cite{dg}. Interestingly, the new
identifications of these solutions with a pair of oppositely-charged extremal black
holes (rather than with a conventional singular point-like dipole) have appeared only afterwards.
To our knowledge, Gross and Perry \cite{gp} were the first to find the Kaluza-Klein 
analog of the Bonnor's solution and identify it with a ``self-gravitating'' 
pole-antipole configuration. Later on, it has been further refined by the work of
Dowker et al. \cite{dowker} where a thorough analysis of the structure of the solutions
and the introduction of the background magnetic field were carried out. Even in these
works, the interpretation was made on the basis of the particular structure of the 
Kaluza-Klein theory, which cannot be applied to solutions with other values of the
dilaton. It is also interesting to note that the Kaluza-Klein dipole solution of
Gross and Perry has lately been uplifted to its counterpart in $d=11$ M-theory
by Sen \cite{sen} which, via the M/IIA string duality, can be interpreted as the
$D6-\bar{D6}$ pair solution \cite{emparan2, hongsu2} in $d=10$ IIA supergravity
theory. That the magnetic dipole solution of Einstein-Maxwell-dilaton theories, including
of course the Bonnor's solution, actually describes a {\it dihole}, i.e., a pair of
static, oppositely-charged extremal black holes with regular horizons was first
demonstrated by Emparan \cite{emparan}. Motivated particularly by this work of Emparan, 
in the present work, we would like to present the construction and extensive analysis
of a solution in the context of the Brans-Dicke-Maxwell theory representing a pair of
static, oppositely-charged extremal black holes. As such, it might be relevant to 
mention the current status of Brans-Dicke (BD) theory as a viable theory of gravity
and describe briefly the strategy to be employed to construct a new solution of
BD-Maxwell theory from a known solution of Einstein-Maxwell theory. \\
Perhaps the Brans-Dicke theory \cite{bd} is the most studied and hence the best-known 
of all the alternative theories of classical gravity to Einstein's general
relativity, . This theory can be thought of as a minimal extension
of general relativity designed to properly accomodate both Mach's principle \cite{weinberg}
and Dirac's large number hypothesis \cite{weinberg}. Namely, the theory employs the
viewpoint in which the Newton's constant $G$ is allowed to vary with space
and time and can be written in terms of a scalar (``BD scalar'') field as
$G = 1/ \Phi $.  Like in Einstein's general relativity, to find the exact solutions 
of the highly non-linear BD field equations is a formidable task. For this reason, algorithms
generating exact, new solutions from the known solutions of simpler situations
either of the BD theory or of the conventional Einstein gravity
have been actively looked for and actually quite a few have been found. 
To the best of our knowledge, methods thus far discovered
along this line includes those of Janis et al., Buchdahl, McIntosh, Tupper \cite{jbmt},
Tiwari and Nayak \cite{tn}, and Singh and Rai \cite{sr}. In particular, Tiwari and Nayak \cite{tn} proposed
an algorithm that allows us to generate stationary, axisymmetric solutions in
vacuum BD theory from the known Kerr solution \cite{kerr} in vacuum Einstein theory and 
later on Singh and Rai generalized this method to the one that generates 
stationary, axisymmetric, charged solutions in BD-Maxwell theory from the
known Kerr-Newman (KN) solution \cite{kerr} in Einstein-Maxwell theory. 
Thus in the present work, we shall employ the solution-generating algorithm of Singh and Rai
in order to construct a Bonnor-type magnetic dipole solution in BD-Maxwell theory from
the known Bonnor solution in Einstein-Maxwell theory and discuss its unfamiliar nature in full 
detail.

\begin{center}
{\rm\bf II. Solution-generating algorithm of Singh and Rai}
\end{center}

We begin by briefly reviewing
the algorithm proposed first by Tiwari and Nayak \cite{tn} and generalized later by Singh 
and Rai \cite{sr}. Consider the BD-Maxwell theory described by the action 
\begin{eqnarray}
S = \int d^4x \sqrt{g}\left[{1\over 16\pi}\left(\Phi R - \omega {{\nabla_{\alpha}\Phi
\nabla^{\alpha}\Phi }\over \Phi}\right) - {1\over 4}F_{\alpha \beta}F^{\alpha \beta}\right]
\end{eqnarray}
where $\Phi $ is the BD scalar field and $\omega $ is the generic parameter of the
theory. Extremizing this action then with respect to the metric $g_{\mu \nu}$, the
BD scalar field $\Phi $, and the Maxwell gauge field $A_{\mu}$ (with the field strength
$F_{\mu \nu}=\nabla_{\mu}A_{\nu}-\nabla_{\nu}A_{\mu}$) yields the classical field
equations given respectively by 
\begin{eqnarray}
G_{\mu \nu} &=& R_{\mu \nu} - {1\over 2}g_{\mu \nu}R = {8\pi \over \Phi}T^{M}_{\mu \nu}
+ 8\pi T^{BD}_{\mu \nu}, \nonumber \\
{\rm where} \\
T^{M}_{\mu \nu} &=& F_{\mu \alpha}F_{\nu}^{\alpha} - {1\over 4}g_{\mu \nu}F_{\alpha \beta}
F^{\alpha \beta}, \nonumber \\
T^{BD}_{\mu \nu} &=& {1\over 8\pi}\left[{\omega \over \Phi^2}(\nabla_{\mu}\Phi \nabla_{\nu}\Phi
- {1\over 2}g_{\mu \nu}\nabla_{\alpha}\Phi \nabla^{\alpha}\Phi) + {1\over \Phi}(\nabla_{\mu}
\nabla_{\nu}\Phi - g_{\mu \nu}\nabla_{\alpha}\nabla^{\alpha}\Phi)\right] \nonumber \\
{\rm and} \nonumber \\
\nabla_{\alpha}\nabla^{\alpha}\Phi &=& {8\pi \over (2\omega + 3)}T^{M\lambda}_{\lambda} = 0,
~~~~\nabla_{\mu}F^{\mu \nu} = 0, ~~~~\nabla_{\mu}\tilde{F}^{\mu \nu} = 0 \nonumber  
\end{eqnarray} 
with the last equation being the Bianchi identity and $\tilde{F}_{\mu \nu}={1\over 2}
\epsilon_{\mu \nu}^{\alpha \beta}F_{\alpha \beta}$. And the Einstein-Maxwell theory is
the $\omega \rightarrow \infty$ limit of this BD-Maxwell theory. As mentioned in the introduction, 
the BD theory
can be thought of as a minimal extension of general relativity designed to properly accomodate
both Mach's principle \cite{weinberg} and Dirac's large number hypothesis \cite{weinberg}.
Namely, the theory employs the viewpoint in which Newton's constant $G$ is allowed to vary with
space and time and can be written in terms of a scalar (``BD scalar'') field as 
$G = 1/\Phi$. Note also that in the action and hence in the
classical field equations, there are no {\it direct} interactions between the BD scalar field
$\Phi$ and the ordinary matter, i.e., the Maxwell gauge field $A_{\mu}$. Indeed this is the 
essential feature of the BD scalar field $\Phi$ that distinguishes it from ``dilaton'' fields
in other scalar-tensor theories such as Kaluza-Klein theories or low-energy effective string
theories where the dilaton-matter couplings generically occur as a result of dimensional reduction.
(Here we would like to stress that we shall work in the context of original BD theory format not
some conformal transformation of it.)
As a matter of fact, it is the original spirit \cite{bd} of BD theory of gravity in which the BD scalar
field $\Phi$ is prescribed to remain strictly massless by forbidding its direct interaction
with matter fields. Now the algorithm of Tiwari and Nayak, and Singh and Rai goes as follows.
Let the metric for a stationary, axisymmetric, charged solution to Einstein-Maxwell field
equations take the form
\begin{eqnarray}
ds^2 = - e^{2U_{E}}(dt + W_{E}d\phi)^2 + e^{2(k_{E}-U_{E})}[(dx^1)^2 + (dx^2)^2] +
h^2_{E}e^{-2U_{E}}d\phi^2 
\end{eqnarray}
while the metric for a stationary, axisymmetric, charged solution to BD-Maxwell field 
equations be
\begin{eqnarray}
ds^2 = - e^{2U_{BD}}(dt + W_{BD}d\phi)^2 + e^{2(k_{BD}-U_{BD})}[(dx^1)^2 + (dx^2)^2] +
h^2_{BD}e^{-2U_{BD}}d\phi^2
\end{eqnarray}
where $U$, $W$, $k$ and $h$ are functions of $x^1$ and $x^2$ only. The significance of the
choice of the metric in this form has been thoroughly discussed by Matzner and Misner \cite{mm}
and Misra and Pandey \cite{mp}. Tiwari and Nayak, and Singh and Rai first wrote down the Einstein-Maxwell
and BD-Maxwell field equations for the choice of metrics in eq.(3) and (4) respectively.
Comparing the two sets of field equations closely, they realized that stationary, axisymmetric
solutions of the BD-Maxwell field equations are obtainable from those of Einstein-Maxwell
field equations provided certain relations between metric functions hold. \\
That is, if $(W_{E}, ~k_{E}, ~U_{E}, ~h_{E}, ~A^{E}_{\mu})$ form a stationary, axisymmetric solution
to the Einstein-Maxwell field equations for the metric in eq.(3), then a corresponding
stationary, axisymmetric solution to the BD-Maxwell field equations for the metric in eq.(4) is
given by $(W_{BD}, ~k_{BD}, ~U_{BD}, ~h_{BD}, ~A^{BD}_{\mu})$ where
\begin{eqnarray}
W_{BD} &=& W_{E}, ~~~k_{BD} = k_{E}, ~~~U_{BD} = U_{E} - {1\over 2}\log \Phi, \\
h_{BD} &=& [h_{E}]^{(2\omega - 1)/(2\omega + 3)}, ~~~\Phi = [h_{E}]^{4/(2\omega + 3)},
~~~A^{BD}_{\mu} = A^{E}_{\mu}. \nonumber
\end{eqnarray}
In the following subsections, we shall apply this solution-generating algorithm to the construction of
the Brans-Dicke gravity Minkowski spacetime (i.e., the empty spacetime solution of the Brans-Dicke
theory) and the Bonnor-type magnetic dipole solution of Brans-Dicke-Maxwell theory.
\\
\\
{\rm\bf (1) Empty spacetime solution in Brans-Dicke theory}
\\
\\
We begin with the construction and interpretation of the empty spacetime solution of
Brans-Dicke theory which will serve as a reference background geometry for all the other spacetime
solutions of Brans-Dicke-Maxwell theory.  Then it will allow us eventually to establish the general 
physical interpretation of all the new solutions that result upon performing the solution-generating 
algorithm presented above to the known solutions in general relativity.
Consider the Minkowski spacetime metric which clearly is a solution to the vacuum Einstein
equation
\begin{eqnarray}
ds^2 = -dt^2 + dr^2 + r^2 d\theta^2 + r^2 \sin^2 \theta d\phi^2.
\end{eqnarray}
For later comparison with its counterpart in Brans-Dicke theory, we also point out that the 
Newton's constant $G$ is indeed a {\it constant} in this Einstein theory. In order to set this
Minkowski metric into the form well prepared for the solution-generating method given above,
we first consider the change of radial coordinate $r = e^{R}$ and hence $dr = e^{R}dR$ which 
leads us to rewrite the eq.(6) into
\begin{eqnarray}
ds^2 = -dt^2 + e^{2R}(dR^2 + d\theta^2) + e^{2R} \sin^2 \theta d\phi^2.
\end{eqnarray}
Then identifying it with the standard form given in eq.(3), we can read off the metric components
as
\begin{eqnarray}
e^{2U_{E}} &=& 1, ~~~W_{E} = 0, \\
e^{2k_{E}} &=& e^{2R}, ~~~h^2_{E} = e^{2R}\sin^2 \theta.  \nonumber
\end{eqnarray}
Now using the solution-generating rule in eq.(5) in the algorithm by Tiwari and Nayak, 
and Singh and Rai, we can now construct the empty spacetime solution of BD theory as
\begin{eqnarray}
ds^2 &=& -\left(r^2 \sin^2 \theta \right)^{-2/(2\omega+3)}dt^2 + \left(r^2 \sin^2 \theta \right)^{2/(2\omega+3)}
[dr^2 + r^2 d\theta^2] \nonumber \\
&+& \left(r^2 \sin^2 \theta \right)^{-2/(2\omega+3)} r^2 \sin^2 \theta d\phi^2,  \\
\Phi (r, \theta) &=& \frac{1}{G}\left(r^2 \sin^2 \theta \right)^{2/(2\omega+3)}  \nonumber
\end{eqnarray}
where we restored the ``bare'' Newton's constant $G$ in the expression for the BD scalar field solution.
Among other things, for $\omega \to \infty$, it reduces to the Minkowski spacetime metric with
$\Phi = 1/G = {\rm constant}$ in eq.(6) as it should since in this limit the Brans-Dicke theory 
goes over to the Einstein's general relativity. \\
Clearly, the solution-generating algorithm of Singh and Rai {\it breaks} the spherical symmetry
along the way from the Minkowski spacetime to the Brans-Dicke gravity empty spacetime as one can
see in eq.(9) above. Close inspection reveals that this happens due to the emergence of
non-trivial BD scalar field which essentially represents the inverse of the spacetime-dependent
effective Newton's constant as mentioned earlier, i.e., $\Phi (r, \theta) = G^{-1}_{eff}(r, \theta)$.
In other words, since the inverse of the BD scalar field representing the strength of the
gravitational interaction changes from point to point, both the spherical symmetry and the
asymptotic flatness of the usual Minkowski spacetime are broken accordingly. \\
We now attempt to provide the physical interpretation of the nature of this empty spacetime
solution of BD theory in a careful manner. This, however, cannot be achieved unless some particular
finite value of the BD parameter is specified. Therefore we first need to appreciate the physical
meaning of the dimensionless BD parameter $\omega$. In the BD gravity action given in eq.(1), the
term $\sim \omega (\nabla_{\alpha}\Phi \nabla^{\alpha}\Phi /\Phi)$, like other terms in the action,
should be finite. Thus large $\omega$ indicates the regime where the BD scalar field $\Phi$ remains
nearly constant as the inverse of the present value of the Newton's constant. Namely, this regime
amounts to the Einstein's general relativity limit. On the other hand, the small $\omega$ indicates
the regime in which the BD scalar field varies sizably with space and time and hence the theory
deviates largely from the general relativity. We have already noted that in the limit $\omega \to \infty$, 
the BD gravity empty spacetime in eq.(9) does reduce to the Minkowski spacetime metric with
$\Phi = 1/G = {\rm constant}$ as it should. Thus next we consider the nature of this spacetime solution
in the other limit, namely for small $\omega$ values. We shall particularly pick the value of the
$\omega$ parameter to be $\omega = -2$ which lies in the range of later interest, 
$-5/2 \leq \omega < -3/2$ (which has particular
significance that will be explained in the next subsection where we shall discuss the Bonnor-type 
magnetic dipole solution in Brans-Dicke-Maxwell theory).
For this value of the BD $\omega$ parameter, the BD gravity empty spacetime solution becomes
\begin{eqnarray}
g_{tt} &=& -(r^2 \sin^2 \theta)^2, ~~~g_{rr} = g_{\theta \theta}/r^2 = \frac{1}{(r^2 \sin^2 \theta)^2},
~~~g_{\phi \phi} = (r^2 \sin^2 \theta)^3,  \nonumber \\
\Phi (r, \theta) &=& \frac{1}{G}(r^2 \sin^2 \theta)^{-2}. 
\end{eqnarray}
Among others, it is interesting to note that from the behavior of the metric function $g_{tt}$, one
can realize that near the origin $r=0$ and the symmetric axis $\theta = 0, \pi$, the {\it infinite
time delay} occurs and hence all the dynamics (if exist) would be frozen there. This is reminiscent
of what happens at the horizon of a black hole. Next, from the behavior of the metric functions 
$g_{rr}$ and $g_{\theta \theta}$, one can notice that both the origin and the symmetry axis are
{\it infinite} proper distance away ! Namely, they turn out to be internal infinities. Again this
is very reminiscent of the feature of a black hole horizon which is indeed an internal infinity
to a distant observer. We now attempt to interpret this peculiar geometric structure of the BD
gravity empty spacetime solution in terms of the {\it spacetime-dependent} behavior of the BD
scalar field  which, as mentioned earlier, serves as the inverse of an 
{\it effective} Newton's constant, $\Phi (r, \theta) \equiv G^{-1}_{eff}(r, \theta)$. First, as we just
pointed out, both the origin and the symmetry axis are {\it infinite} proper distance away
\begin{eqnarray}
\int^{0}_{r} \sqrt{g_{rr}}dr &=& \int^{0}_{r} \frac{1}{r^2 \sin^2 \theta}dr \to \infty, \\
\int^{0}_{\theta} \sqrt{g_{\theta \theta}}d\theta &=& \int^{0}_{\theta} \frac{1}{r \sin^2 \theta}d\theta \to \infty
\end{eqnarray}
while the asymptotic region $r = \infty$ is {\it finite} proper distance away
\begin{eqnarray}
\int^{\infty}_{r} \sqrt{g_{rr}}dr &=& \int^{\infty}_{r} \frac{1}{r^2 \sin^2 \theta}dr < \infty .
\end{eqnarray}
It seems that this can be attributed to the spacetime-dependent behavior of the {\it effective}
Newton's constant
\begin{eqnarray}
\Phi^{-1}(r, \theta) = G_{eff}(r, \theta) = G (r^2 \sin^2 \theta)^2.
\end{eqnarray}
Namely, since the Newton's constant is a measure of the strength of the gravitational interaction,
this spacetime-dependent behavior of $G_{eff}$ implies that the gravitational interaction vanishes
at the origin and along the symmetry axis while it diverges asymptotically as $r\to \infty$.
To be a little more concrete, the regions such as the origin and the symmetry axis are essentially
inaccessible as the gravitational interaction is effectively absent and hence no test object
would get accelerated (or at least be in motion) there. Meanwhile the coordinate infinity $r=\infty$
can be in a finite proper distance as the gravitational interaction grows as $\sim r^4$
and eventually diverges at $r=\infty$ and thus any test particle there would get accelerated infinitely.
Namely, the physical (proper) distance between the two spacetime points gets larger as the
effective gravitational interaction there gets weaker whereas it gets smaller as the effective
interaction there gets stronger. In order to support this type of interpretation in terms of the
spacetime-dependent behavior of the effective Newton's constant, one can take some other values of
the BD $\omega$ parameter and see if essentially the same interpretation still holds true.
For instance if one takes, say, $\omega =1/2$ or $5/2$ which lie outside of the above-mentioned
range, $-5/2\leq \omega <-3/2$, just the other way around is the case.
Namely, the origin and the symmetry axis are finite proper distance away (indeed instantly accessible)
as the gravitational interaction strength diverges there while the asymptotic region $r=\infty$
is infinite proper distance away as the gravitational interaction strength vanishes there.
And of course in this analysis, the ``backreaction'' of the
test particle to the background spacetime that would modify the background spacetime structure
itself is neglected. The peculiar features of the BD gravity empty spacetime geometry, particularly
the fact that the symmetry axis actually is an internal infinity, in turn, leads to the 
occurrence of rather an embarrassing conical singularity and we now turn to this issue.
Notice that the conical singularity, or more specifically, the deficit angle can be defined in the
following manner. Take the ratio between the proper circumference and the proper radius of a small
limiting circle around the symmetry axis $\theta = 0, \pi$. If this ratio comes out to be exactly
$2\pi$, then there is no deficit angle. If instead it turns out to be less than $2\pi$, then
there exists non-vanishing deficit angle. Being guided by this criterion, we now compute the
possible deficit angle in the BD gravity empty spacetime geometry.
\begin{eqnarray}
\delta_{(0,\pi)} = 2\pi - \arrowvert {\Delta \phi
d\sqrt{g_{\phi\phi}} \over \sqrt{g_{\theta\theta}}d\theta}
\arrowvert_{\theta =0, \pi} = 2\pi
\end{eqnarray}
where $\Delta \phi$ is the period of the azimuthal angle coordinate that will be chosen as $2\pi$
and we used $g_{\theta \theta} = r^2/(r^2 \sin^2 \theta)^2$ and
$g_{\phi \phi} = (r^2 \sin^2 \theta)^3$ and hence 
$(g_{\theta\theta})^{-1/2}d\sqrt{g_{\phi\phi}}/d\theta |_{\theta=0, \pi} = 0$. 
This demonstrates that this seemingly pathological ``$2\pi$-angle deficit'' is indeed a direct
consequence of our earlier observation that the proper radius of a small limiting circle
around the symmetry axis diverges, $\int^{0}_{\theta} \sqrt{g_{\theta \theta}}d\theta \to \infty $
as the symmetry axis can be thought of as an internal infinity. Put differently, 
``$2\pi$-angle deficit'' is {\it not} a generic pathology but just another manifestation of the 
emergence of the internal infinity, i.e., the infinitely far symmetry axis.
\\
\\
{\rm\bf (2) Bonnor-type magnetic dipole solution in Brans-Dicke-Maxwell theory}
\\
\\
Next, we apply the same method to obtain the Bonnor-type magnetic dipole solution in 
BD-Maxwell theory from the known Bonnor solution \cite{bonnor} in Einstein-Maxwell theory. 
And to do so, again one needs some preparation which involves casting the Bonnor solution given in 
Boyer-Lindquist coordinates $(t, r, \theta, \phi)$ in the metric form in eq.(3) by performing a 
coordinate transformation (of $r$ alone) suggested by Misra and Pandey \cite{mp}. 
Namely, we start with Bonnor's magnetic dipole (dihole) 
solution of Einstein-Maxwell theory written in Boyer-Lindquist coordinates \cite{bonnor, emparan}
\begin{eqnarray}
ds^2 &=& \left(1 - \frac{2Mr}{\Sigma}\right)^{2}
\left[-dt^2 + \frac{\Sigma^4}{(\Delta+(M^2+a^2)\sin^2 \theta)^3}({dr^2\over \Delta} + d\theta^2)\right] \nonumber \\
&+& \left(1 - \frac{2Mr}{\Sigma}\right)^{-2}\Delta \sin^2 \theta d\phi^2,  \\
A &=& A_{\mu}dx^{\mu} = \frac{2aMr\sin^{2} \theta}{\Delta + a^2 \sin^2 \theta}d\phi \nonumber
\end{eqnarray}
where $\Sigma = r^2-a^2 \cos^2 \theta $ and $\Delta = r^2 -2Mr - a^2$ with $M$ denoting the ADM mass 
and $a$ representing (roughly) the proper separation between the poles \cite{emparan}.
Consider now the transformation of the radial coordinate 
\begin{eqnarray}
r = e^{R} + M + {(M^2+a^2)\over 4}e^{-R}
\end{eqnarray}
which gives $dr^2/ \Delta = dR^2$. Indeed, this transformation has been deduced from the associated
one introduced by Misra and Pandey \cite{mp} for the case of single rotating (Kerr) black hole.
Then Bonnor's magnetic dipole solution can now be cast in the form in eq.(3),
i.e.,
\begin{eqnarray}
ds^2 &=& -\left(1 - \frac{2ML}{\Sigma}\right)^{2}dt^2
+ \frac{\Sigma^4}{(\Delta+(M^2+a^2)\sin^2 \theta)^3}\left(1 - \frac{2ML}{\Sigma}\right)^{2}
(dR^2 + d\theta^2) \nonumber \\
&+& \left(1 - \frac{2ML}{\Sigma}\right)^{-2}\Delta \sin^2 \theta d\phi^2
\end{eqnarray}
with now $\Sigma = L^2-a^2 \cos^2 \theta $ and $\Delta = L^2 -2ML - a^2$ where we set, as a
short-hand notation, $L \equiv e^{R} + M + {(M^2+a^2)\over 4}e^{-R}$.
Then identifying it with the standard form given in eq.(3), we can read off the metric components as
\begin{eqnarray}
e^{2U_{E}} &=& \left(1 - \frac{2ML}{\Sigma}\right)^2 = 
\left({\Delta + a^2\sin^2 \theta \over \Sigma}\right)^2,
~~~W_{E} = 0, \\
e^{2k_{E}} &=& \frac{\Sigma^4}{(\Delta + (M^2+a^2)\sin^2 \theta)^3}\left(1 - \frac{2ML}{\Sigma}\right)^4,
~~~h^2_{E} = \Delta \sin^2 \theta.  \nonumber
\end{eqnarray}
Now using the solution-generating rule in eq.(5) in the algorithm by Tiwari and Nayak, 
and Singh and Rai, we can now construct the Bonnor-type magnetic dipole solution in BD-Maxwell theory as
\begin{eqnarray}
ds^2 &=& - \left(1 - \frac{2ML}{\Sigma}\right)^2
(L^2 - 2ML - a^2)^{-2/(2\omega+3)}\sin^{-4/(2\omega+3)}\theta dt^2 \nonumber \\
&+& \frac{\Sigma^4}{(\Delta + (M^2+a^2)\sin^2 \theta)^3}\left(1 - \frac{2ML}{\Sigma}\right)^2
(L^2 - 2ML - a^2)^{2/(2\omega+3)}\sin^{4/(2\omega+3)}\theta [dR^2 + d\theta^2] \nonumber \\
&+& \left(1 - \frac{2ML}{\Sigma}\right)^{-2}(L^2 - 2ML - a^2)^{(2\omega+1)/(2\omega+3)}\sin^{2(2\omega+1)/(2\omega+3)}\theta 
d\phi^2, \nonumber \\
\Phi (R, \theta) &=& (L^2 - 2ML - a^2)^{2/(2\omega+3)}\sin^{4/(2\omega+3)}\theta, \\
A &=& \frac{2aML\sin^2 \theta}{\Delta + a^2 \sin^2 \theta}d\phi. \nonumber
\end{eqnarray}
Finally by transforming back to the standard Boyer-Lindquist coordinates using eq.(7), we arrive at
the Bonnor-type dipole solution in Boyer-Lindquist coordinates \cite{bl} given by
\begin{eqnarray}
ds^2 &=&  \left(\frac{\Delta + a^2 \sin^2 \theta}{\Sigma}\right)^2 \left[ - 
(\Delta \sin^2 \theta)^{-2/(2\omega+3)}dt^2 + (\Delta \sin^2 \theta)^{2/(2\omega+3)}
\frac{\Sigma^4}{(\Delta + (M^2+a^2)\sin^2 \theta)^3} \right. \nonumber \\
&\times & \left. \left(\frac{dr^2}{\Delta} + d\theta^2\right)\right] +
\left(\frac{\Sigma}{\Delta + a^2 \sin^2 \theta}\right)^2(\Delta \sin^2 \theta)^{(2\omega+1)/(2\omega+3)}
d\phi^2, \nonumber  \\
\Phi (r, \theta) &=& \frac{1}{G}(\Delta \sin^2 \theta)^{2/(2\omega+3)},
~~~A = \frac{2aMr\sin^2 \theta}{\Delta + a^2 \sin^2 \theta}d\phi 
\end{eqnarray}
where we restored the ``bare'' Newton's constant $G$ in the expression for the BD scalar field solution.
This solution is static and axisymmetric. It is interesting to note that this solution becomes the
flat Minkowski spacetime asymptotically as $r\to \infty$ or for $M=0$ and $a=0$ only as $\omega \to \infty$
which amounts to the general relativity limit. 
And for finite $\omega$, upon setting $M = 0 = a$ as well as $A = 0$, this solution reduces to the
empty spacetime solution of BD theory given in eq.(9) as it should. 
Besides, this solution possesses singularities
along the symmetry axis $\theta = 0,~\pi$ as well as at $r=r_{+}=M+\sqrt{M^2 + a^2}$ 
where $\Delta = 0$ for finite $\omega$. 
As we shall see later on, these are not curvature singularities but {\it coordinate} ones on which 
all the curvature invariants remain finite and indeed the symmetry axis of the geometry represented 
by this solution consists of the semi-infinite lines $\theta = 0,~\pi$ and the segment $r=r_{+}$. 
And as we shall see in a moment, at each of the ``poles'' 
$(r=r_{+},~\theta=0)$, $(r=r_{+},~\theta=\pi)$, lies two extremal 
oppositely charged black holes of the BD-Maxwell theory. 
In addition, from the asymptotic ($r\to \infty$) behavior of the magnetic vector
potential given above, one can deduce that the solution involves the magnetic dipole moment of
$2Ma$. Thus the solution can be identified with a magnetic dipole solution. 
Associated with this last point, note that changing the sign of $a$ amounts to
reversing the orientation of the dipole, so we will consider, without any loss of generality,
$a\geq 0$. Another point worthy of note is that even if we take the limit $a\to 0$, 
this Bonnor-type magnetic dipole solution fails to
reduce to the single Brans-Dicke-Schwarzschild black hole solution. Noticing that the parameter,
$a$, can be regarded (for large $a$) as indicating the proper separation between the two opposite
charges \cite{sen, emparan, hongsu2}, this observation reveals a peculiar feature that as $a\to 0$, the pair of 
opposite magnetic charges (or black holes) never appear to merge and the limiting solution maintains 
axisymmetric structure. This point is reminiscent of the magnetic black dihole solutions in
other Einstein-Maxwell-dilaton theories like Kaluza-Klein or low energy string theories as well as 
the general relativity coupled to the Maxwell theory \cite{emparan}. \\
Originally, Bonnor's magnetic dipole solution in Einstein-Maxwell theory was thought to
describe a singular, point-like dipole. Recently, however, Emparan \cite{emparan} demonstrated
that it actually describes a ``black dihole'' which has regular horizons. Emparan's argument for
the black dihole interpretation of original Bonnor's solution goes as follows. Bonnor's solution
is clearly asymptotically flat as $r\to \infty$ and in this asymptotic region, the gauge field
solution becomes that of a magnetic dipole. Next, although the solution exhibits apparent
singularities at $r=r_{+}$ (where $\Delta = r^2-2Mr-a^2 = 0$), they need to be treated more
carefully. Note first that the axis of symmetry of the solution (namely the fixed point set of the
Killing field $(\partial/\partial \phi)$) consists of the semi-infinite lines, $\theta = 0, \pi$
(running from $r=r_{+}$ to $r= \infty$) and the segment $r=r_{+}$ (running from $\theta = 0$ to
$\theta = \pi$). Then a crucial feature of the Bonnor's solution is that at each of the poles,
$(r=r_{+}, \theta = 0)$ and $(r=r_{+}, \theta = \pi)$, lies a distorted extremal 
Reissner-Nordstrom black holes. \\
Thus it is of equal interest to study if the Bonnor-type dipole solution in BD-Maxwell theory 
constructed in the present work can describe dihole geometry as well and if so, under what condition.
Among others, the bottomline condition for possible dihole interpretation concerns the particular
values of the BD $\omega$ parameter. Note that some time ago it has been pointed
out \cite{hongsu} that in order for the BD-Reissner-Nordstrom solution (or more generally, for the 
BD-Kerr-Newman solution) to describe a non-trivial, {\it regular}, charged black hole spacetime, the
$\omega$-parameter of the BD theory should take values in the range, $-5/2\leq \omega <-3/2$. 
Here, by ``regular'' it means that the black hole spacetime possesses regular Killing horizons
at which the invariant curvature polynomials such as $R$, $R_{\mu\nu}R^{\mu\nu}$, and
$R_{\mu\nu\alpha\beta}R^{\mu\nu\alpha\beta}$ remain nonsingular. Thus hereafter we shall assume
that the BD $\omega$ parameter takes values in this particular range. \\
Again, our eventual objective is the dihole interpretation of the BD-Bonnor solution in eq.(21).
Prior to this, however, it seems necessary to discuss the peculiar geometric structure of this
BD-Bonnor solution that appears to inherit essentially the same generic features of the BD gravity 
empty spacetime solution we studied earlier in the previous subsection. 
Firstly for finite $\omega$ value, this BD-Bonnor solution in eq.(21) fails to be asymptotically
flat for the following reason. Namely, the solution-generating algorithm of Singh and Rai
{\it breaks} the asymptotic flatness along the way essentially due to the emergence of non-trivial
BD scalar field representing the inverse of the spacetime-dependent effective Newton's constant.
This is a generic nature of the Singh and Rai's solution-generating algorithm that we first realized
in the construction of the BD gravity empty spacetime solution discussed in the previous subsection.
Next, the symmetry axis again turns out to be some kind of internal infinity. To see this, consider
the BD-Bonnor metric solution for the value $\omega = -2$ which lies in the range, $-5/2\leq \omega <-3/2$,
\begin{eqnarray}
g_{tt} &\sim& -(\Delta \sin^2 \theta)^2, ~~~g_{rr} = g_{\theta \theta}/\Delta \sim \frac{1}{(\Delta \sin^2 \theta)^2},
~~~g_{\phi \phi} \sim (\Delta \sin^2 \theta)^3,  \nonumber \\
\Phi (r, \theta) &=& \frac{1}{G}(\Delta \sin^2 \theta)^{-2}. 
\end{eqnarray} 
From this, we can readily realize that the symmetry axis is {\it infinite} proper distance away
\begin{eqnarray}
\int^{0}_{\theta} \sqrt{g_{\theta \theta}}d\theta \sim \int^{0}_{\theta} \frac{1}{\sin^2 \theta}d\theta \to \infty
\end{eqnarray}
while the asymptotic region $r = \infty$ is {\it finite} proper distance away
\begin{eqnarray}
\int^{\infty}_{r} \sqrt{g_{rr}}dr \sim \int^{\infty}_{r} \frac{1}{\Delta \sin^2 \theta}dr < \infty .
\end{eqnarray}
Like before, this can be attributed to the spacetime-dependent behavior of the {\it effective}
Newton's constant
\begin{eqnarray}
\Phi^{-1}(r, \theta) = G_{eff}(r, \theta) = G (\Delta \sin^2 \theta)^2.
\end{eqnarray}
Namely, since the Newton's constant is a measure of the strength of the gravitational interaction,
this spacetime-dependent behavior of $G_{eff}$ implies that the gravitational interaction vanishes
on the symmetry axis while it diverges asymptotically as $r\to \infty$.
More specifically, the region such as the symmetry axis is essentially
inaccessible as the gravitational interaction is effectively absent there whereas 
the coordinate infinity $r=\infty$ can be in a finite proper distance as the gravitational 
interaction grows as $\sim r^4$ and eventually diverges at $r=\infty$. 
That is, the physical (proper) distance between the two spacetime points gets larger as the
effective gravitational interaction there gets weaker whereas it gets smaller as the effective
interaction there gets stronger. Again, this is indeed the same generic feature of any
spacetime solution of the BD gravity theory that we first encountered in the case of the BD gravity
empty spacetime solution before.  Then next, we are led to anticipate that the internal infinity
nature of the symmetry axis we just observed would likewise lead to the occurrence of the
conical singularity in the geometry described by this BD-Bonnor solution.
Thus we turn now to the study of possible conical singularity
structure of this BD-Bonnor solution. First observe that the
rotational Killing field $\psi^{\mu} = (\partial/\partial
\phi)^{\mu}$ possesses vanishing norm, i.e., $\psi^{\mu}\psi_{\mu}
= g_{\alpha \beta}\psi^{\alpha}\psi^{\beta} = g_{\phi\phi} =
\left[\Sigma /(\Delta + a^2 \sin^2 \theta )\right]^2
(\Delta \sin^2 \theta)^{(2\omega+1)/(2\omega+3)}= 0$ at
the locus of $r=r_{+}$ as well as along the semi-infinite lines
$\theta=0, \pi$ for $-5/2\leq \omega <-3/2$. 
This implies that $r=r_{+}$ can be thought of as
a part of the symmetry axis of the solution. Namely unlike the
other familiar axisymmetric solutions, for the case of the
BD-Bonnor solution under consideration (and of course for the original 
Bonnor solution as well), the endpoints of the
two semi-axes $\theta =0$ and $\theta =\pi$ do not come to join at
a common point. Instead, the axis of symmetry is completed by the
segment $r=r_{+}$. And as $\theta$ varies from $0$ to $\pi$, one
moves along the segment from $(r=r_{+}, \theta=0)$ where a black hole is
situated to $(r=r_{+}, \theta=\pi)$ where the other hole with opposite charge is placed.
Obviously, then the natural question to be addressed is whether or not the
conical singularities arise on different portions of the symmetry
axis. Thus once again we define the notion of conical angle deficit (or excess). 
Consider that ; if $C$ is the proper length of the circumference around the
symmetry axis and $R$ is its proper radius, then the occurrence of
a conical angle deficit (or excess) $\delta$ would manifest itself
if $(dC/dR)|_{R\rightarrow 0}=2\pi - \delta$ \cite{emparan}. We now proceed to
evaluate this conical deficit (or excess). 
The conical angle deficit along the axes $\theta =0, \pi$
and along the segment $r=r_{+}$ are given respectively by 
\begin{eqnarray}
\delta_{(0,\pi)} &=& 2\pi - \arrowvert {\Delta \phi
d\sqrt{g_{\phi\phi}} \over \sqrt{g_{\theta\theta}}d\theta}
\arrowvert_{\theta =0, \pi} = 2\pi, \\
\delta_{(r=r_{+})} &=& 2\pi - \arrowvert {\Delta \phi
d\sqrt{g_{\phi\phi}} \over \sqrt{g_{rr}}dr} \arrowvert_{r=r_{+}} =
2\pi  \nonumber
\end{eqnarray}
where, of course, we used the BD-Bonnor metric solution given above in eq.(21) for
$\omega$-parameter of the BD theory taking values in the range, $-5/2\leq \omega <-3/2$.
Just as we expected, we encounter the seemingly pathological ``$2\pi$-angle deficit'' again.
This, of course, is no surprise as it is not a generic pathology but just another manifestation of
the emergence of the internal infinity, i.e., the symmetry axis. Namely, as the symmetry axis
is infinite proper distance away, we, as a consequence, have
$(g_{\theta\theta})^{-1/2}d\sqrt{g_{\phi\phi}}/d\theta |_{\theta=0, \pi} = 0$ which, in turn, 
results in $\delta_{(0,\pi)} = 2\pi$. Once again, this conical singularity structure is not
a separate pathology but a generic feature that appears to inherit essentially the same thing 
of the BD gravity empty spacetime solution we studied earlier in the previous subsection. \\
Note that these peculiar features are the generic characteristics of the static solutions of
the BD-Maxwell theory that appear as long as we 
confine our interest to the case of the black ``dihole'' interpretation, namely, as long as 
we assume $-5/2\leq \omega <-3/2$ and are really somethings that do not show up in the case of the
Einstein-Maxwell theory or other scalar-tensor theories such as the minimal Kaluza-Klein theory 
or the bosonic sector of the low energy effective string theories \cite{emparan}. \\
Lastly, therefore, we need to ask, prior to its black dihole interpretation, whether or not
the BD-Bonnor solution given in eq.(21) is qualified at all to be endowed with any meaning as 
a physical spacetime solution. To answer this, recall that for a value of the BD $\omega$ parameter
in the range of interest $-5/2\leq \omega <-3/2$, the BD gravity empty spacetime solution was
given an acceptable physical meaning or interpretation in terms of the spacetime-dependent
behavior of the effective Newton's constant. Next, since we keep employing the same 
solution-generating algorithm of Singh and Rai, the Bonnor-type solution of BD-Maxwell theory
can be thought of as the one that results when one puts a pair of dipole (or oppositely-charged
black holes as we shall demonstrate shortly) on opposite sides along the symmetry axis of the empty
spacetime. As can be expected, however, the generic peculiar features of the spacetime such as 
the internal infinity nature of the symmetry axis and its consequence, namely the conical 
singularity still survive the addition process of the dipole content. Thus the BD-Bonnor solution
deserves essentially the same physical meaning or interpretation as that of the BD gravity
empty spacetime solution we studied earlier. The only delicate issue to be addressed would be whether
or not the dipole lying along the symmetry axis is really physically meaningful when the symmetry
axis is an internal infinity. At least it seems to us that it should nevertheless be of enough
physical interest as long as we take the BD gravity empty spacetime solution seriously as the one
that embodies some spirit of Mach's principle, namely that the Newton's constant may take different
values in different spacetime points. This issue also can be viewed from a different angle.
Here we are worried about whether or not a pair of black holes lying along an infinitely far symmetry 
axis can be of any meaning. Along this line we need to recall an essentially
analogous (single) black hole case in which the horizon (namely, the black hole itself) is infinite 
proper distance away to an asymptotic observer. In this sense, the BD-Bonnor spacetime is indeed
nearly as weird as the familiar black hole solutions. \\
Since we accepted the BD-Bonnor solution as having a physical meaning,
we now proceed to elaborate on the magnetic black dihole spacetime interpretation of this solution 
in some more detail. Particulary, we would like to show explicitly that at each of the poles,
$(r=r_{+}, \theta = 0)$ and $(r=r_{+}, \theta = \pi)$, lies a distorted extremal 
BD-Reissner-Nordstrom black holes in a manner similar to what happens in the Bonnor's magnetic dihole
solution in Einstein-Maxwell theory first demonstrated by Emparan.  To this end, we begin by 
performing the change of coordinates from $(r, \theta)$ to $(\rho, \bar{\theta})$ given by the 
following transformation law \cite{sen, emparan, hongsu2}
\begin{eqnarray}
r = r_{+} + \frac{\rho}{2}(1 + \cos \bar{\theta}), 
~~~\sin^2 \theta = \frac{\rho}{\sqrt{M^2 + a^2}}(1 - \cos \bar{\theta}) 
\end{eqnarray}
where $\Delta(r_{+})=0$, namely $r_{+}=M+\sqrt{M^2+a^2}$ as introduced earlier.
Then one can realize that upon changing the coordinates as given above and taking $\rho$ to be
much smaller than any other length scale involved so as to get near each pole, the Bonnor-type
magnetic dipole solution in BD-Maxwell theory given earlier becomes
\begin{eqnarray}
ds^2 &\simeq& g^2(\bar{\theta})\left[-(\rho^2 \sin^2 \bar{\theta})^{-2/(2\omega +3)}
\left(\frac{\rho}{q}\right)^2 dt^2 + (\rho^2 \sin^2 \bar{\theta})^{2/(2\omega +3)}\left(\frac{q}{\rho}\right)^2
(d\rho^2 + \rho^2 d\bar{\theta}^2)\right] \nonumber \\
&+& g^{-2}(\bar{\theta})(\rho^2 \sin^2 \bar{\theta})^{-2/(2\omega +3)}\left(\frac{q}{\rho}\right)^2
\rho^2 \sin^2 \bar{\theta}d\phi^2, \nonumber \\
\Phi(\rho, \bar{\theta}) &\simeq& \frac{1}{G}(\rho^2 \sin^2 \bar{\theta})^{2/(2\omega +3)}, \\
A &\simeq& q\frac{a}{\sqrt{M^2 + a^2}}g^{-1}(\bar{\theta})(1 - \cos \bar{\theta})d\phi \nonumber
\end{eqnarray}
where $q \equiv \frac{Mr_{+}}{\sqrt{M^2 + a^2}}$ and
$g(\bar{\theta}) \equiv [\cos^2(\frac{\bar{\theta}}{2}) + 
\frac{a^2}{(M^2 + a^2)}\sin^2(\frac{\bar{\theta}}{2})].$
Certainly, this limit of the solution can be identified with the ``near-horizon limit''
$(r\to r_{+} ~{\rm or} ~\rho \to 0)$ of a {\it distorted} extremal charged black hole in
BD-Maxwell theory with the distortion being described by the factor $g(\bar{\theta})$. And due to
this distortion factor $g(\bar{\theta})$, the geometry is not spherically symmetric, rather it is
elongated along the axis in a prolate shape. And of course this deformation of the horizon geometry
(to a prolate spheroid) is due to the field created by the other hole and possibly by the conical
defect to which we shall come back in a moment. \\
The limiting geometry above were valid for arbitrary value of ``$a$'' as long as we remain close
enough to (one of the) poles. If instead we consider the limit of very large ``$a$'', while
keeping $(r-r_{+})$ and $a \sin^2 \theta$ finite, the Bonnor-type magnetic dipole solution
reduces, in this time, to
\begin{eqnarray}
ds^2 &\simeq& -\left(1 + \frac{q}{\rho}\right)^{-2}(\rho^2 \sin^2 \bar{\theta})^{-2/(2\omega +3)}dt^2 
\nonumber \\
&+& \left(1 + \frac{q}{\rho}\right)^{2}\left[(\rho^2 \sin^2 \bar{\theta})^{2/(2\omega +3)}
(d\rho^2 + \rho^2 d\bar{\theta}^2) + (\rho^2 \sin^2 \bar{\theta})^{-2/(2\omega +3)}
\rho^2 \sin^2 \bar{\theta}d\phi^2 \right],  \nonumber \\
\Phi(\rho, \bar{\theta}) &\simeq& \frac{1}{G}(\rho^2 \sin^2 \bar{\theta})^{2/(2\omega +3)}, \\
A &\simeq& q(1 - \cos \bar{\theta})d\phi \nonumber
\end{eqnarray}
with $q = \frac{Mr_{+}}{\sqrt{M^2 + a^2}}\to M$ as $a\to \infty$. \\
As we shall demonstrate below, this limit of the solution can be recognized as representing the
extremal BD-Reissner-Nordstrom black hole (i.e., the charged single black hole solution in 
BD-Maxwell theory). Indeed, this result was rather expected since physically, taking the limit
$a\to \infty$, amounts to pushing one of the poles to a large distance and studying the geometry
of the remaining pole. We now demonstrate
in an explicit manner that the above limit of the BD-Bonnor solution indeed represents the
BD-Reissner-Nordstrom black hole spacetime. To this end, we consider the change of the radial
coordinate (which takes $\rho$ to the standard Boyer-Lindquist radial coordinate $R$) given by
\begin{eqnarray}
R = (\rho + q).
\end{eqnarray}
Then one ends up with
\begin{eqnarray}
ds^2 &=& \left[R^2\left(1-\frac{q}{R}\right)^2 \sin^2 \bar{\theta}\right]^{-2/(2\omega +3)}
\left[- \left(1-\frac{q}{R}\right)^2 dt^2 + R^2 \sin^2 \bar{\theta}d\phi^2\right] \\
&+& \left[R^2\left(1-\frac{q}{R}\right)^2 \sin^2 \bar{\theta}\right]^{2/(2\omega +3)}
\left[-\left(1-\frac{q}{R}\right)^{-2} dR^2 + R^2 d\bar{\theta}^2 \right]. \nonumber 
\end{eqnarray}
Certainly, this can be identified with the extremal, charged non-rotating black hole solution in
BD-Maxwell theory found in \cite{hongsu}. To check this, one only needs to take the non-rotating
limit of the BD-Kerr-Newman black hole solution there and set $r\to R$, $\theta \to \bar{\theta}$
and $e\to q$ (see the appendix). 
Then one gets the BD-Reissner-Nordstrom black hole solution which is, in our
present coordinates $(t, R, \bar{\theta}, \phi)$, given by eq.(31).  \\
To summarize, based on these observations, we may conclude that the BD-Bonnor solution given
above indeed describes a {\it black dihole} geometry. \\  

\begin{center}
{\rm\bf III. Bonnor-type magnetic dipole solution with Melvin-type universe content
in Brans-Dicke-Maxwell theory}
\end{center}

In the preceding section, we have constructed the Bonnor-type magnetic dipole solution in 
BD-Maxwell theory and demonstrated that indeed it can be identified with a ``black dihole''
solution particularly for values of the BD $\omega$-parameter in the range $-5/2\leq \omega <-3/2$
representing a configuration in which a pair of oppositely charged extreme regular black holes is 
placed at each of the poles, $(r=r_{+}, \theta=0)$ and $(r=r_{+}, \theta=\pi)$ respectively.
However, since we failed to address the detailed nature of conical singularities along the symmetry 
axis, $\theta = 0, \pi$ and along the segment, $r=r_{+}$ for the reason stated in detail earlier,
we are still away from the satisfactory
understanding of how the two oppositely charged black holes can be sustained in an (un)stable 
equilibrium against the combined gravitational (same mass) and gauge or electromagnetic (opposite
charge) attractions. Usually, if there are {\it finite} conical angle deficits along the semi-infinite axes,
$\theta = 0, \pi$ (as is the case with generalized Bonnor solutions in Einstein-Maxwell-dilaton
theories \cite{emparan}), one interprets these angle deficits as the presence of {\it cosmic strings}
providing tension that pulls the pole and the antipole at the endpoints apart. Namely, in this
interpretation, the tension generated by the cosmic strings counterbalances the combined
gravitational and gauge attractions and holds the dipole (or dihole) apart aginst the collapse.
Indeed, the emergence of the conical singularities signals the instability of the configuration and by 
the recourse to the presence of cosmic strings with proper tension, one achieves the stability
as the line $r=r_{+}$, $0<\theta <\pi$ joining the poles can be made to be completely non-singular
then. Alternatively, perhaps another very relevant approach toward the stabilization of the system 
one can turn to would be to introduce an external magnetic field, aligned with the axis joining the
dipole, to counterbalance the combined gravitational and gauge attractive forces by pulling them
apart. By properly ``tuning'' the strength of the magnetic field, the internal stresses along the
axis would be rendered to vanish. And this can be checked in an explicit manner if the introduction
of the magnetic field with ``right'' strength leads to the vanishing of the finite conical singularities
along the symmetry axis, $\theta = 0, \pi$ and along the segment, $r=r_{+}$ \cite{hongsu2, emparan} at the
same time.  Particularly for the case at hand in which the cosmic string interpretation is not 
available for the reason stated earlier, this second option involving the introduction of 
external magnetic field looks quite appropriate. 
Note also that the conical singularity structure of the generalized Bonnor solution
in Einstein-Maxwell-dilaton theories \cite{emparan} and its cure via the introduction of external
magnetic field is reminiscent of the Ernst's prescription \cite{ernst} for the elimination of 
conical singularities of the charged C-metric in Einstein-Maxwell theory. In this section, therefore,
we attempt to construct the Bonnor-type magnetic dipole solution of Brans-Dicke-Maxwell theory
containing the magnetic field content that asymptotes to the ``Melvin-type'' magnetic universe
which is a flux tube that provides the best possible approximation to a self-gravitating uniform
magnetic field. And to this end, we shall employ the Ehlers-Harrison-type transformation generalized
to the BD-Maxwell theory which, in the Einstein-Maxwell theory, has been known to generate an
axisymmetric solution with a magnetic field content from a known axisymmetric solution without.
It, however, is interesting to note that there indeed are two different ways of achieving this.
The first route is to start with the Bonnor-type magnetic dipole solution obtained in the preceding
section and apply to it the generalized Ehlers-Harrison transformation to arrive at our
destination, i.e., the BD-Bonnor solution with magnetic Melvin-type universe content.
Meanwhile, the second route is to start instead with the Bonnor solution with magnetic Melvin
universe content in Einstein-Maxwell theory (which is already known in the literature \cite{emparan}) 
obtained from the original Bonnor solution via the Ehlers-Harrisn transformation and then
apply to it the Singh and Rai's algorithm which is known as a solution-generating method that allows
one to generate a stationary axisymmetric charged solutions in BD-Maxwell theory from their
counterparts in Einstein-Maxwell theory. As we shall see in a moment, these two methods yield the
same final solution and this indicates, among others, that the two separate solution-generating 
algorithms, the Singh and Rai's one and the Ehlers-Harrison transformation indeed commute. 
And in the following we shall discuss both two routes. \\

{\rm \bf (1) Route 1}  \\

In this subsection, therefore, we construct the BD-Bonnor solution with magnetic Melvin-type 
universe content via the first route stated above. 
Consider the BD gravity theory coupled to Maxwell field described by the action
\begin{eqnarray}
S = \int d^4x \sqrt{g}\left[{1\over 16\pi}\left\{\Phi R - \omega g^{\mu\nu}{{\partial_{\mu}\Phi
\partial_{\nu}\Phi }\over \Phi}\right\} - {1\over 4}g^{\mu\alpha}g^{\nu\beta}
F_{\mu \nu}F_{\alpha \beta}\right].
\end{eqnarray}
We now consider the Weyl-rescaling given by
\begin{eqnarray}
g_{\mu\nu} &=& \Omega^2(x)\tilde{g}_{\mu\nu}, ~~~\Phi = M^2_{pl}e^{\Psi/\Psi_{0}} \\
{\rm with} ~~~\Omega^2(x) &=& \frac{M^2_{pl}}{\Phi}, ~~~\Psi^2_{0} = \frac{M^2_{pl}}{16\pi}(2\omega+3)
\nonumber
\end{eqnarray}
where $M_{pl}$ denotes the Planck mass.
Under this Weyl-rescaling, the action above transforms to
\begin{eqnarray}
\tilde{S} = \int d^4x \sqrt{\tilde{g}}\left[{1\over 16\pi}\left\{\tilde{R} - 
\frac{1}{2} \tilde{g}^{\mu\nu}{\partial_{\mu}\Psi \partial_{\nu}\Psi }\right\} - 
{1\over 4}\tilde{g}^{\mu\alpha}\tilde{g}^{\nu\beta}F_{\mu \nu}F_{\alpha \beta}\right].
\end{eqnarray}
Namely, upon this Weyl-rescaling, the action of the BD gravity coupled to the Maxwell field takes the 
form of that of the Einstein-Maxwell-dilaton theory in which the dilaton field
is a minimally-coupled, massless (Weyl-rescaled) BD scalar field. \\
Then we are now ready to perform the ``generalized'' Ehlers-Harrison transformation 
(already known in the literature \cite{emparan}) which takes an axisymmetric solution
of the Einstein-Maxwell-dilaton theory to another such solution containing particularly the
(asymptotically) Melvin-type magnetic universe content. It is given by
\begin{eqnarray}
\tilde{g}^{'}_{ij} &=& \lambda^2 \tilde{g}_{ij} ~~(i,j\neq \phi),
~~~\tilde{g}^{'}_{\phi\phi} = \lambda^{-2} \tilde{g}_{\phi\phi}, \\
\Psi^{'} &=& \Psi, 
~~~A^{'}_{\phi} = -\frac{2}{B\lambda}[1 + \frac{1}{2}BA_{\phi}] + \frac{2}{B} \nonumber \\
{\rm with} ~~~\lambda &\equiv& [1 + \frac{1}{2}BA_{\phi}]^2 + \frac{1}{4}B^2 \tilde{g}_{\phi\phi}.
\nonumber 
\end{eqnarray}
Indeed, the virtue of the above Weyl-rescaling on the action of BD-Maxwell theory was to 
make use of the generalized Ehlers-Harrison transformation already known in the literature.
Since we achieved this goal, now the remaining task is to translate this generalized
Ehlers-Harrison transformation given in terms of the Weyl-rescaled fields
$(\tilde{g}_{\mu\nu}, A_{\mu}, \Psi)$ back into that given in the original fields
$(g_{\mu\nu}, A_{\mu}, \Phi)$, which now reads
\begin{eqnarray}
g^{'}_{ij} &=& \lambda^2 g_{ij} ~~(i,j\neq \phi),
~~~g^{'}_{\phi\phi} = \lambda^{-2} g_{\phi\phi}, \\
\Phi^{'} &=& \Phi, 
~~~A^{'}_{\phi} = -\frac{2}{B\lambda}[1 + \frac{1}{2}BA_{\phi}] + \frac{2}{B} \nonumber \\
{\rm with} ~~~\lambda &\equiv& [1 + \frac{1}{2}BA_{\phi}]^2 + \frac{1}{4}B^2 \frac{\Phi}{M^2_{pl}}
g_{\phi\phi}.
\nonumber 
\end{eqnarray}
Finally, upon performing this generalized Ehlers-Harrison transformation (22) on the BD-Bonnor solution
given earlier in eq.(11), one gets the BD-Bonnor solution with magnetic Melvin-type 
universe content which is given by
\begin{eqnarray}
ds^2 &=& \Lambda^2 \left[ - 
(\Delta \sin^2 \theta)^{-2/(2\omega+3)}dt^2 + (\Delta \sin^2 \theta)^{2/(2\omega+3)}
\frac{\Sigma^4}{(\Delta + (M^2+a^2)\sin^2 \theta)^3}
\left(\frac{dr^2}{\Delta} + d\theta^2\right)\right] \nonumber \\
&+& \Lambda^{-2}(\Delta \sin^2 \theta)^{(2\omega+1)/(2\omega+3)}
d\phi^2, \nonumber  \\
\Phi (r, \theta) &=& \frac{1}{G}(\Delta \sin^2 \theta)^{2/(2\omega+3)},  \\
A &=& \frac{1}{2\Lambda \Sigma}\left\{4Mra + B[(r^2-a^2)^2 + \Delta a^2 \sin^2 \theta ]\right\}
d\phi, \nonumber \\
{\rm with} ~~~\Lambda &\equiv& \left(\frac{\Delta + a^2 \sin^2 \theta}{\Sigma}\right)\lambda \nonumber \\
&=& \frac{1}{\Sigma}\left\{(\Delta + a^2 \sin^2 \theta) + 2BaMr\sin^2 \theta + \frac{1}{4}
B^2\sin^2 \theta [(r^2-a^2)^2 + \Delta a^2 \sin^2 \theta ]\right\}.  \nonumber
\end{eqnarray}

{\rm \bf (2) Route 2}  \\

Next, we attempt to construct the BD-Bonnor solution with magnetic Melvin-type 
universe content via the second route described before. And to this end, we start with the Bonnor
solution with magnetic Melvin universe content in Einstein-Maxwell theory discussed in \cite{emparan}
and then apply to it the Singh and Rai's algorithm. Again in Boyer-Lndquist-type coordinates,
the Bonnor solution involving Melvin's magnetic universe content is given by
\begin{eqnarray}
ds^2 &=& \Lambda^2 \left[ -dt^2 + \frac{\Sigma^4}{(\Delta + (M^2+a^2)\sin^2 \theta)^3}
\left(\frac{dr^2}{\Delta} + d\theta^2\right)\right] 
+ \Lambda^{-2}\Delta \sin^2 \theta d\phi^2, \nonumber  \\
A &=& \frac{1}{2\Lambda \Sigma}\left\{4Mra + B[(r^2-a^2)^2 + \Delta a^2 \sin^2 \theta ]\right\}
d\phi, \\
{\rm with} ~~~\Lambda &=& \frac{1}{\Sigma}
\left\{(\Delta + a^2 \sin^2 \theta) + 2BaMr\sin^2 \theta + \frac{1}{4}
B^2\sin^2 \theta [(r^2-a^2)^2 + \Delta a^2 \sin^2 \theta ]\right\}.  \nonumber
\end{eqnarray}
From this point on, the procedure to generate the BD-Bonnor solution with magnetic Melvin-type 
universe content by applying the Singh and Rai's algorithm to this solution is quite straightforward
and essentially the same as that to generate the BD-Bonnor solution from the Bonnor solution in
Einstein-Maxwell theory we discussed earlier in section II. To briefly sketch the procedure,
consider, again, the coordinate transformation given in eq.(17) which gives
$dr^2/\Delta = dR^2$. Then the Bonnor solution with magnetic Melvin universe content above can be
cast into the form in eq.(3) from which we can read off
\begin{eqnarray}
e^{2U_{E}} &=& \Lambda^2, ~~~W_{E} = 0, \\
e^{2k_{E}} &=& \frac{\Sigma^4}{(\Delta + (M^2+a^2)\sin^2 \theta)^3}\Lambda^4,
~~~h^2_{E} = \Delta \sin^2 \theta. \nonumber
\end{eqnarray}
Then using the solution-generating rule in eq.(5) in the algorithm by Tiwari and Nayak, 
and Singh and Rai, and then transforming back to the standard Boyer-Lindquist coordinates using eq.(17),
finally we can construct the BD-Bonnor solution with magnetic Melvin-type
universe content which turns out to be precisely the same as the one obtained earlier via the
``route 1'', namely eq.(37).   \\
Now let us appreciate the meaning of this agreement once again. If we recall the nature of these two
methods, the ``route 1'' involves Singh and Rai's solution-generating method followed by the
Ehlers-Harrison transformation while the ``route 2'' involves actions of the reversed order.
This indicates, among others, that the two separate solution-generating methods, that of Singh and Rai
and that of Ehlers-Harrison indeed commute. 
Although this realization may not be so surprising, at least the lesson we learned from it is the
fact that the Ehlers-Harrison-type transformation we derived in eq.(36) in the context of the
BD-Maxwell theory was indeed correct and hence can be applied to other cases of interest.  \\
Now upon successfully constructing the BD-Bonnor solution with magnetic Melvin-type 
universe content, we now turn to the question asking if this solution can describe dihole geometry 
as well. In order to address this issue, we, as before, perform the change of
coordinates from $(r, \theta)$ to $(\rho, \bar{\theta})$ given in eq.(27) and examine the solution
for very small value of $\rho$ to get
\begin{eqnarray}
ds^2 &\simeq& \tilde{g}^2(\bar{\theta})\left[-(\rho^2 \sin^2 \bar{\theta})^{-2/(2\omega +3)}
\left(\frac{\rho}{q}\right)^2 dt^2 + (\rho^2 \sin^2 \bar{\theta})^{2/(2\omega +3)}\left(\frac{q}{\rho}\right)^2
(d\rho^2 + \rho^2 d\bar{\theta}^2)\right] \nonumber \\
&+& \tilde{g}^{-2}(\bar{\theta})(\rho^2 \sin^2 \bar{\theta})^{-2/(2\omega +3)}\left(\frac{q}{\rho}\right)^2
\rho^2 \sin^2 \bar{\theta}d\phi^2, \nonumber \\
\Phi(\rho, \bar{\theta}) &\simeq& \frac{1}{G}(\rho^2 \sin^2 \bar{\theta})^{2/(2\omega +3)}, \\
A &\simeq& q\left[\frac{a}{\sqrt{M^2 + a^2}}+Bq\right]\tilde{g}^{-1}(\bar{\theta})(1 - \cos \bar{\theta})d\phi \nonumber
\end{eqnarray}
where $q \equiv \frac{Mr_{+}}{\sqrt{M^2 + a^2}}$ and
$\tilde{g}(\bar{\theta}) \equiv [\cos^2(\frac{\bar{\theta}}{2}) + 
(\frac{a}{\sqrt{M^2 + a^2}} + Bq)^2 \sin^2(\frac{\bar{\theta}}{2})]$.
And lastly, the magnetic charge of the solution is 
\begin{eqnarray}
Q_{m} = q\left[\frac{a}{\sqrt{M^2 + a^2}}+Bq\right]^{-1}\left(\frac{\Delta \phi}{2\pi}\right)
\end{eqnarray}
where $\Delta \phi$ denotes the yet unspecified period of the azimuthal angle $\phi$.
It is the ``physical charge'' of a single pole computed using Gauss' law
$Q_{m} = \frac{1}{4\pi}\int_{S^2}F_{[2]}$ with $S^2$ being any topological sphere surrounding
each pole and $F_{[2]}=dA$. Again, this limit of the solution can be identified with the 
``near-horizon limit''
$(r\to r_{+} ~{\rm or} ~\rho \to 0)$ of a {\it distorted} extremal charged black hole in
BD-Maxwell theory with the distortion now being described by the factor $\tilde{g}(\bar{\theta})$. 
And the horizon, i.e., the $\rho = 0$ surface lacks spherical symmetry, instead, it is a prolate
spheroid. \\
Lastly, we turn to the conical singularity
structure of this BD-Bonnor solution with magnetic Melvin-type universe content. 
Using the metric solution given above in eq.(37) for
$\omega$-parameter of the BD theory taking values in the range, $-5/2\leq \omega <-3/2$,
the conical angle deficit along the axes $\theta =0, \pi$ 
and along the segment $r=r_{+}$ are given respectively by 
\begin{eqnarray}
\delta_{(0,\pi)} &=& 2\pi - \arrowvert {\Delta \phi
d\sqrt{g_{\phi\phi}} \over \sqrt{g_{\theta\theta}}d\theta}
\arrowvert_{\theta =0, \pi} = 2\pi, \\
\delta_{(r=r_{+})} &=& 2\pi - \arrowvert {\Delta \phi
d\sqrt{g_{\phi\phi}} \over \sqrt{g_{rr}}dr} \arrowvert_{r=r_{+}} =
2\pi.  \nonumber
\end{eqnarray}
To our dismay, we are still left with the same embarrassing results as before when the magnetic
field content was absent. Once again, this seemingly pathological result, namely the ``$2\pi$-angle
deficits'' along both the semi-infinite lines $\theta=0, \pi$ and the segment $r=r_{+}$ can be
attributed to the fact that the metric structure of eq.(37) leading to the emergence of the
``coordinate singularities'' along $\theta=0, \pi$ and $r=r_{+}$ essentially remained the same
(even if we have introduced the external magnetic field content into the BD-Bonnor solution) as the
one of eq.(21) which, as stressed several times, exhibits the internal infinity nature of the
symmetric axis.
Earlier, we mentioned that particularly for the case at hand in which the cosmic string 
interpretation is not available, perhaps this second option involving the introduction of 
external magnetic field could be quite an appropriate approach toward the stabilization of 
the dihole system. And this can be checked in an explicit manner if the introduction
of the magnetic field with proper strength leads to the vanishing of the conical singularities
along the symmetry axis, $\theta = 0, \pi$ and along the segment, $r=r_{+}$ at the
same time. Apparently, however, this hope of ours failed again.
As a result, any attempt to remove the conical singularities along the symmetry axis 
and along the segment, $r=r_{+}$ in terms of the cosmic string interpretation or by introducing
an external (Melvin-type) magnetic field of proper strength failed in the present case of the
Bonnor-type spacetime solution in BD-Maxwell theory. And we know that this failure can essentially 
be attributed to the internal infinity nature of the symmetry axis to begin with. 
Nevertheless, we anticipate that, although
it cannot be checked in an explicit fashion for the reason stated above, the introduction of
external magnetic field, aligned with the axis joining the dihole, can {\it in principle} 
counterbalance the combined 
gravitational and gauge attractive forces and hence eventually stabilize the configuration.  

\begin{center}
{\rm\bf IV. Summary and Discussions}
\end{center}

In the present work, the construction and extensive analysis
of a solution in the context of the BD-Maxwell theory representing a pair of
static, oppositely-charged extremal black holes have been performed.
To this end, the algorithm of Singh and Rai's is employed which is known to generate
stationary, axisymmetric, charged solutions in BD-Maxwell theory from the
known such solutions in Einstein-Maxwell theory.
Indeed it was originally thought that Bonnor's solution in Einstein-Maxwell theory describes 
a singular point-like magnetic dipole. Lately, however, it has been demonstrated \cite{emparan} 
that it instead
may describe a black {\it dihole}. Thus it is of equal interest to examine if the Bonnor-type 
dipole solution in BD-Maxwell theory constructed in this work can describe dihole 
geometry as well. Particularly for values of the BD $\omega$-parameter in the range 
$-5/2\leq \omega <-3/2$, it has been demonstrated that indeed it can be identified with a black dihole
configuration. The supporting argument for this conclusion is that although both the BD gravity
empty spacetime and the BD-Bonnor solution exhibit some peculiar geometric structures such as the
internal infinity nature of the symmetry axis, they should not be viewed as a pathology but instead 
should be thought of as the manifestation of the Mach's principle that the BD theory itself attempts 
to embody.  Then followed the discussion of the nature of this new solution particularly concerning its 
stability issue. Obviously, the configuration described by this black dihole solution involves an
instability arising from the combined gravitational (same mass) and electromagnetic (opposite charge)
attractions and it indeed is represented by the emergence of conical singularities. 
As we realized in the text, however, any attempt to address the nature of possible
conical singularity structure of the BD-Bonnor solution along the symmetry axis, $\theta = 0, \pi$ 
and along the segment, $r=r_{+}$ failed essentially due to the internal infinity nature of the
symmetry axis and the coordinate singularity nature at $r=r_{+}$ as long as we 
confined our interest to the case of the black dihole interpretation, namely, as long as 
$-5/2\leq \omega <-3/2$. As a result, the cosmic string interpretation of the conical singularities
was not available and therefore we turned to the second option involving the introduction of 
the Melvin's magnetic universe content, i.e., the external magnetic field. Along the way, we also
noticed that actually there are two possible ways of achieving this and the ``route 1'' involves 
Singh and Rai's solution-generating method followed by the Ehlers-Harrison transformation while 
the ``route 2'' involves actions of the reversed order. It, then, has been realized that the two
methods yield the same final solution which indicates that the two separate solution-generating 
methods, that of Singh and Rai and that of Ehlers-Harrison indeed commute. Unfortunately, however,
this second approach toward the stabilization of the dihole system failed again as the metric structure 
in eq.(37) leading to the emergence of the ``coordinate singularities'' along 
$\theta=0, \pi$ and $r=r_{+}$ essentially remained the same even if we have introduced the 
external magnetic field content into the BD-Bonnor solution. This failure manifests itself
since one cannot determine for the case at hand the proper strength of the magnetic field that can
eliminate the conical singularities along the symmetry axis, $\theta = 0, \pi$ and along the segment, 
$r=r_{+}$ at the same time.  Despite this technical difficulties, however, it may be anticipated
that {\it in principle} the introduction of external magnetic field, aligned with the axis joining 
the dihole, can counterbalance the combined gravitational and electromagnetic attractive forces and 
eventually stabilize the configuration. Lastly, since the reinterpretation of Bonnor's solution
in Einstein-Maxwell theory as a magnetic black dihole configuration, its generalizations in other
Einstein-Maxwell-dilaton theories such as the Kaluza-Klein or the low-energy string 
theories have been discussed in the literature \cite{emparan}. Thus the only remaining viable
scalar-tensor theory of gravity (coupled to Maxwell theory) left untouched along this line was the
Brans-Dicke-Maxwell theory. And precisely it is this issue that we would like to address in the
present work.

\vspace*{1cm}

\begin{center}
{\rm\bf Appendix : Regular black hole solutions in Brans-Dicke-Maxwell theory}
\end{center}

In this appendix, we shall briefly review the characteristics of the non-trivial, regular
black hole solutions of the BD-Maxwell theory constructed and studied in \cite{hongsu}
and particularly discuss its BD-Reissner-Nordstrom black hole solution limit which is of
some relevance to the present work. \\
The Brans-Dicke-Kerr-Newman (BDKN) solution of the BD-Maxwell theory in Boyer-Lindquist coordinates
is given by \cite{hongsu}
\begin{eqnarray}
ds^2 &=&  (\Delta \sin^2 \theta)^{-2/(2\omega+3)}\left[ - \left({\Delta - a^2\sin^2 \theta \over
\Sigma}\right)dt^2 - {2a\sin^2 \theta (r^2+a^2-\Delta)\over \Sigma}dt d\phi \right. \nonumber \\
&+&\left. \left({(r^2+a^2)^2 - \Delta a^2\sin^2 \theta \over \Sigma }\right)\sin^2 \theta d\phi^2 \right]
+ (\Delta \sin^2 \theta)^{2/(2\omega+3)} \left[ {\Sigma \over \Delta}dr^2 + \Sigma d\theta^2 \right], 
\nonumber \\
\Phi (r, \theta) &=& \frac{1}{G}(\Delta \sin^2 \theta)^{2/(2\omega+3)},
~~~A_{\mu} = -{er\over \Sigma}[\delta^{t}_{\mu} - a\sin^2 \theta \delta^{\phi}_{\mu}] 
\end{eqnarray}
where $\Sigma = r^2+a^2 \cos^2 \theta $ and $\Delta = r^2 -2Mr + a^2 + e^2$ with $M$, $a$, and $e$
denoting the ADM mass, angular momentum per unit mass, and the electric charge respectively.
Notice that the quantities $\Sigma$ and $\Delta$ above in eq.(43) are defined differently from
those with the same names appeared in the BD-Bonnor solution in eq.(21) in the text. Indeed, they
are related by $a\to ia$. Just like the way how the BD-Bonnor solution has been derived in the
text of the present work, this BDKN solution in BD-Maxwell theory has been constructed starting from 
the Kerr-Newman (KN) solution in Einstein-Maxwell theory and then applying to it the Singh and Rai's 
solution-generating algorithm as well. Note also that the BDKN solution above has possible coordinate 
singularities not only at the outer event horizon where $\Delta = 0$ but also along the symmetry 
axis $\theta = 0, ~\pi$. Thus in order to explore the nature of this singularity along the 
symmetry axis, the computation of invariant curvature polynomials is necessary. 
And it is a straightforward matter to realize that the two invariant curvature polynomials
$R_{\mu\nu}R^{\mu\nu}$ and $R_{\mu\nu\alpha\beta}R^{\mu\nu\alpha\beta}$ become finite both 
on the horizon candidate at which $\Delta = 0$ and along the symmetry axis $\theta = 0, ~\pi$ 
provided the generic BD $\omega$-parameter takes values in the range 
$-5/2 \leq \omega < -3/2$ \cite{hongsu}. \\
Particularly note that if one takes the non-rotating
limit of this BDKN black hole solution by setting $a$ (i.e., the angular momentum per unit mass)
to zero one ends up with the BD-Reissner-Nordstrom black hole solution
\begin{eqnarray}
ds^2 &=& \left[r^2\left(1-\frac{e}{r}\right)^2 \sin^2 \theta\right]^{-2/(2\omega +3)}
\left[- \left(1-\frac{e}{r}\right)^2 dt^2 + r^2 \sin^2 \theta d\phi^2\right] \\
&+& \left[r^2\left(1-\frac{e}{r}\right)^2 \sin^2 \theta \right]^{2/(2\omega +3)}
\left[-\left(1-\frac{e}{r}\right)^{-2} dr^2 + r^2 d\theta ^2 \right] \nonumber 
\end{eqnarray}
which, upon setting $r\to R$, $\theta \to \bar{\theta}$ and $e\to q$, coincides with eq.(31)
in the text. \\
Now it seems relevant to explore thermodynamics and causal structure of this non-trivial BDKN black hole solution 
in eq.(43) in some more detail. Firstly, these BDKN black hole solutions have vanishing surface
gravity at the event horizon, $\kappa_{+}=0$ and hence {\it zero} Hawking temperature, 
$T_{H}=\kappa_{+}/2\pi=0$ provided $-5/2 \leq \omega < -3/2$. In other words, they do not 
radiate and hence are completely ``dark and cold''. Certainly, this is a very bizzare feature in
sharp contrast to evaporating black holes in general relativity. Next, we turn to their causal structure.
The two Killing horizons, i.e., the outer event horizon and the inner Cauchy horizon
turn out to occur precisely at the same locations (i.e., same coordinate distances)
as those of KN black hole solutions in Einstein-Maxwell
theory, i.e., at $r_{\pm} = M \pm (M^2-a^2-e^2)^{1/2}$. Also it is interesting to note that the proper
area of the event horizon at $r = r_{+}$,
\begin{eqnarray}
A = \int_{r_{+}} d\theta d\phi (g_{\theta \theta}g_{\phi \phi})^{1/2} = 4\pi (r^2_{+}+a^2)
\end{eqnarray}
is again exactly the same as that of standard KN black hole spacetime. In addition, its angular velocity 
at the event horizon coincides with that of standard KN solution as well 
\begin{eqnarray}
-W^{-1}_{BD}(r_{+}) = {a\over {r^2_{+}+a^2}} = -W^{-1}_{E}(r_{+}).
\end{eqnarray}
Next, observe that the norm of the time translational Killing field
\begin{eqnarray}
\xi^{\mu}\xi_{\mu} = g_{tt} = - (\Delta \sin^2 \theta)^{-2/(2\omega+3)}
\left[{\Delta - a^2\sin^2 \theta \over \Sigma}\right]
\end{eqnarray}
goes like negative $(r_{-}<r<r_{+})$ $\rightarrow$ positive $(r_{+}<r<r_{s})$ $\rightarrow$ negative
$(r>r_{s})$ with $r_{s} = M + (M^2-a^2\cos^2 \theta - e^2)^{1/2} > r_{+}$ being the larger root of
$\xi^{\mu}\xi_{\mu}$, indicating that $\xi^{\mu}$ behaves as timelike $\rightarrow$ spacelike
$\rightarrow$ timelike correspondingly. And particularly the region in which $\xi^{\mu}$ stays
spacelike extends outside hole's event horizon. This region is the so-called ``ergoregion'' and
its outer boundary on which $\xi^{\mu}$ becomes null, i.e., $r=r_{s}$ is called ``static limit''
since inside of which no observer can possibly remain static. Thus if we recall the location of the
static limit in standard KN black hole solution, we can realize that even the locations of
ergoregions in two black hole spacetimes, KN and BDKN, are the same as well. Namely in two theories,
i.e., the BD-Maxwell theory and the Einstein-Maxwell theory, rotating, charged black hole solutions
turn out to possess {\it identical} causal structure (i.e., the locations of ring singularities,
two Killing horizons and static limits are the same) and hence exhibit the same global topology.
Thus actually what distinguishes the BDKN black hole spacetime from its general relativity's
counterpart, i.e., the KN black hole is the local geometry alone such as the curvature characterized by
the specific $\omega $-values, $-5/2 \leq \omega < -3/2$. \\
Next, we would like to comment on the divergent behavior of the BD scalar field solution on the 
horizon which, in addition to the null Hawking radiation mentioned earlier, is another peculiar 
feature of the solution.  
As we mentioned earlier, the BD theory is an alternative theory to Einstein gravity and the BD 
scalar field represents spacetime-varying effective Newton's constant, not a matter. 
Thus the divergent behavior of the BD scalar field in BD theory essentially represents the
vanishing effective Newton's constant in a certain region of spacetime. Besides, since the energy
density of the BD scalar field $T^{BD}_{\mu\nu}\xi^{\mu}\xi^{\nu}$ vanishes and
hence satisfies the weak energy condition on the horizon at which $\Delta = 0$ (of course for
$-5/2 \leq \omega <-3/2$) \cite{hongsu}, we do not worry too much about the divergent behavior of 
the BD scalar field there. For more detailed account of the nature of the BDKN solution, we refer
the reader to \cite{hongsu}.

\vspace*{1cm}

\begin{center}
{\rm\bf Acknowledgements}
\end{center}

H.Kim was financially supported by the BK21 Project of the Korean Government and
H.M.Lee was supported by the Korean Research Foundation Grant No. D00268 in 2001.

\end{document}